\documentclass[12pt,a4paper]{article}
\usepackage{graphicx}
\usepackage{adjustbox}
\usepackage{multirow}
\usepackage[margin=2cm]{geometry}
\usepackage{amsmath}
\usepackage{amsfonts}
\usepackage{amsthm}
\usepackage{amssymb}
\usepackage{multirow}
\usepackage{caption}
\usepackage{subcaption}

\title {Control charts in a multi product environment: An application study}
\author{ Thomas Muehlenstaedt\\ W. L. Gore \& Associates, 85639 Putzbrunn, Germany }

\begin{document}             

\maketitle
\section*{Abstract}
In this article an SPC case study is presented. 
It consists of monitoring a manufacturing process used for  different products of similar kind.
So far, each of these products is monitored individually. However, if there is e.g. a quality problem with one joint component, this signal will be distributed over different control charts. Therefor a standardization is introduced, similar to short run SPC methods.
Then control chart methods are applied to detect process deviations.  
If there are observations out of control, a next step is taken and process information are analysed in order to derive possible root causes. 


\textbf{Keywords:} machine learning, partition tree, standardization, short run SPC, statistical process control

\section{Introduction}\label{sec_intro}


Detecting an out of control status of a production process is an important problem in industry to assure that products are made with a similar quality over time and to avoid unexpected scrap. In this context a real case study of statistical process control is presented, which is based on data collected at one division of W. L. Gore \& Associates.
On the product portfolio produced in one plant, there is usually a very wide spread of products in a given category.
For all the products produced in one plant, there are some product properties, which are measured throughout the portfolio, that is, for each production lot produced, no matter from which product, this property is evaluated in the lab.
For Gore's Fabrics division, this could be  a measure of breathability for a laminate, i.e. how much moisture can evaporate through a textile composite under fixed environmental conditions (called 'breathability' in the following). This breathability is a key feature of Gore-Tex$^\text{\textregistered}$ laminates. A garment made of a highly breathable laminate will enable a user to perform an activity longer over time by preventing the wearer from overheating but at the same time to protect the wearer from bad weather.

As the products considered here are using different raw materials, different kinds of product constructions and different processes/tooling, this results in very different values for the considered property. Besides the different raw materials, there are other factors, which highly impact the breathability, like e.g. different toolings, machine settings or different operator teams. 

As an example, think of a fire fighter jacket laminate, where the priority is to ensure fire protection properties and hence breathability might not be as high as for a laminate for running wear, where a high breathability is a priority. What is a clear fail in breathability for the running wear laminate is acceptable for the fire fighter laminate in terms of breathability.

A standard approach for control charting could then be to set up a control chart for each product separately, see (\cite{WheelerChambersSPC2010}, \cite{MontgomeryIntroSPC2001}) for introductions into the topic of statistical process control.  Control limits would be calculated from a phase of a stable process and new measurement values would be compared to the control limits in order to judge if the process is still stable. 

In general this is a recommendable procedure. However, in the data situation at hand, a failure might take long to correctly be identified as an error and potentially it takes even longer to find the root cause of the failure. Consider e.g. a raw material being used in several products. If this raw material has a quality problem it will lead to failures not just in one finished product but in several. A separated analysis of these products might not detect the issue directly. More attractive is to have a procedure, which analyses the data jointly so that a bigger chance exists to correctly detect the unstable process and identify the root cause. Also looking at a high number of control charts on a regular basis (e.g. weekly) is not really acceptable. Another problem is that for many products only a limited number of production lots are produced over e.g. one year, which leads to the situation that for these products standard control charting techniques like individual range charts are not meaningful.

Therefore a procedure is suggested here, consisting of 3 steps:
	\begin{enumerate}
		\item Normalize the lab measurements per product according to the product average and variability.
		\item On the normalized measurements perform a control chart, e.g. IR charts or exponentially weighted moving average charts.
		\item If an unstable process is detected, use data mining techniques to identify potential root causes.
	\end{enumerate}
The first and the second step are mostly in line with conventional control chart methods and are going in a similar direction as short run SPC methods, see e.g. \cite{Griffith1996SPCLongShortRuns}, chapter 3, where observations of different products are transformed in several ways to set up joint control charts over different products. However, short run SPC is developed for cases, where there are either not enough observations available for calculating control limits and/or for cases, where only a few different products are combined for statistical process controll.
The third step is an attempt to go a step further and to use existing explanatory data to identify root causes.

This work is motivated  by a number of papers by Prof. Steiner from the University of Waterloo ( \cite{Steiner2000MonitoringSurgicalOutcomesBiostats}, \cite{Woodall2015MonitoringAndImprovementOfSurgicalOutcome},    \cite{Steiner2014MonitoringRiskAdjustedMedicalOutcomesAllowingForChangesOverTime} and several others, see publication list of Professor Steiner). The overall theme there is to do process control for medical surgeries. For the task to do process control for the performance of surgeons over time or between different surgeons, the different risk profiles of the patients needs to be considered. Hence a risk adjustment is introduced such that the outcomes of surgeries can be combined in a meaningful way in a control chart. Although the context in this article is completely different the basic questions is similar, as in both situations the original data available needs to be transformed so that a joint plotting in a control chart is meaningful.

In the following the steps of this procedure are explained. In chapter 2 the used data are further explained. In chapter 3 the methods used are presented followed by an application study. The article is concluded by a discussion and summary.

\section{Data description}
The data used and described here are real data sets from Gore. They have been rescaled and transformed for intellectual property reasons. The range of the data is over one year in order to have a sufficient range of observations.
The observations of the lab measurement results are noted by $y_{ij} \in \mathbb{R}$, where the index $i$ refers to different products and the index $j$ refers to different production lots of this product.
These data can be assumed to be normally distributed: $y_{ij} \sim \mathbb{N}(\mu_i, \sigma_i^2)$.
It is assumed, that one measurement per production lot of every product is done.

\begin{figure}[ht]
	\centering
	\includegraphics[width=\textwidth]{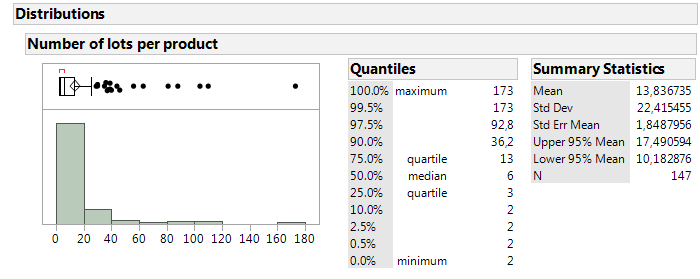}
	\caption{Number of lots per product in the example data set.}
	\label{fig:7DistNoOfLots}
\end{figure}

\begin{figure}[ht]
  \centering
  \includegraphics[width=\textwidth]{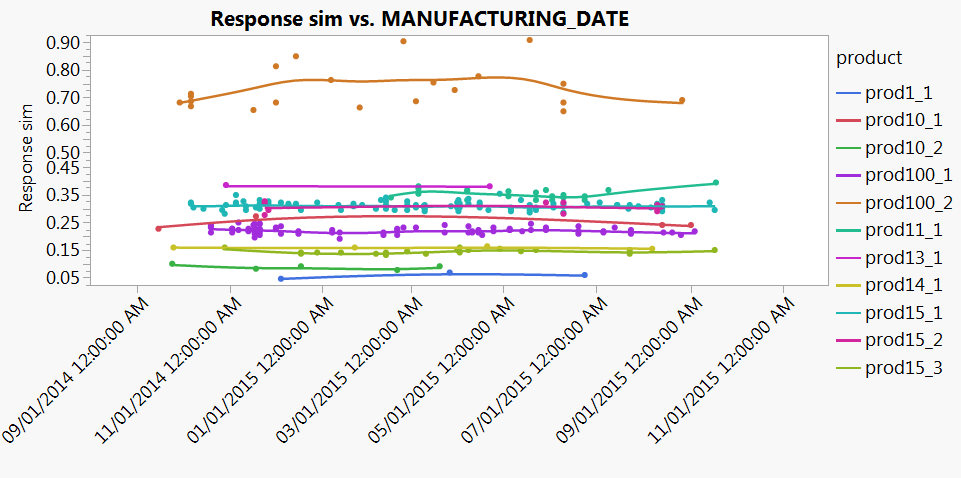}
\caption{Simulated breathability results over time by product, unscaled.}
\label{fig:1WVPoverTime}
\end{figure}

In the data set used, there are $i = 1, \dots, 147$ products, with very low number of production lots from 2 or 3 up to $n_i = 173$ production lots over one year. In Figure \ref{fig:7DistNoOfLots}, a histogram of the number of lots per product is given in order to provide information about a typical data set in this application. As can be seen, the majority of all products has less than 10 lots, hence a solid application of conventional control charting would only be possible for a small subset of products.
Also, in order to clarify why looking at products separately in a control chart might not give a realistic picture of the underlying process, Figure \ref{fig:1WVPoverTime} is included.
Here the data for a small subset of products are plotted over a range of 1 year against the date of production.
As products are produced in a somehow random manner, the data per product might only give a small fraction of the information of the process.\\
For each observation $y_{ij}$ there is a corresponding vector $x_{ij} \in \mathbb{R}^K$, describing conditions of that production lot. These information include raw material types for that specific product, process information, e.g. which process line was used, which tooling configuration was used, timing information like manufacturing date and shift, and lab information as for example lab station used and lab operator who performed the test. 
Depending on the context, this could include all types of information, which potentially help to identify possible root causes for unstable processes. 
One important note to keep in mind is that these data are not planned according to a designed experiment but that they are observational data. So conclusion drawn from any data mining out of these observations can suffer from confounding, different powers, partial confounding and unclearity of cause and effect. 

\begin{figure}[ht]
  \centering
  \includegraphics[width=\textwidth]{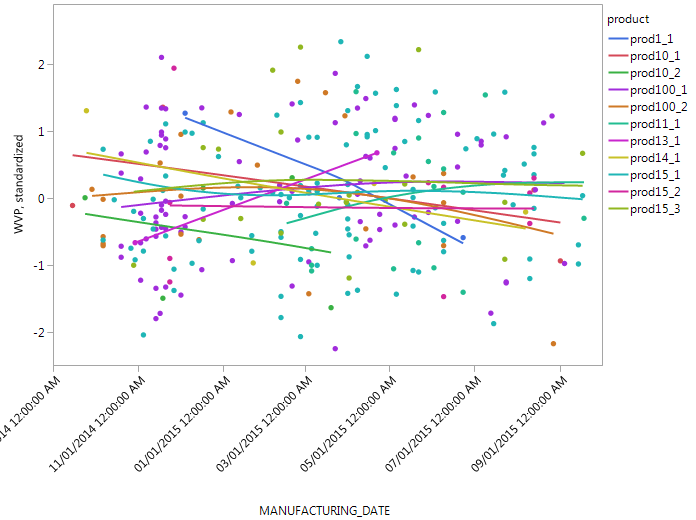}
\caption{Simulated breathability results over time by product, scaled.}
\label{fig:2WVPoverTimeScaled}
\end{figure}
\section{Methods}
\subsection{Standardizing}


The first step of data processing to have a unified view on the production data is to standardize them. 
A first obvious way is to standardize by mean and standard deviation:
\begin{equation} \label{eq:1Standardizing}
	y^{(s)}_{ij} = \frac{y_{ij} - \bar{y}_i}{s_i},
\end{equation}
with $\bar{y}_i$ being the observed mean per product and $s_i$ being the observed standard deviation. Similar transformations are also proposed in the context of short run SPC methods, as described in \cite{Griffith1996SPCLongShortRuns}. In \cite{Quesenberry1991Qchart} and \cite{CastilloMontgomery1994QchartEnahncement}, even a next step is taken and the data are transformed also taking into account the sample size of different sample sizes. 
As the data used tend to produce outliers during the measurement process, a robust version of the standardization can be used as well:
\begin{equation} \label{eq:2Standardizingrobust}
	y^{(r)}_{ij} = \frac{y_{ij} - \tilde{y}_i}{s_i^{(r)}},
\end{equation}
with $\tilde{y}_i$ being a robust estimator of the average per product and $s_i^{(r)}$ being a robust estimator for the variability of the product.
Options for standardizing are the median or a trimmed version of the average, where we will focus here on the median. For the variability the median absolute deviation (MAD), the interquartile range (IQR) or a robust version of the standard deviation (r std dev) could be used. The robust standard deviation is based on an M-estimator as described in \cite{HuberRonchettiRobustStats2009}, chapter 5.


After discussing control charting in the following section, the different options for average and variability estimators will be compared by simulations in order to derive recommendations. 
	
Having standardized the values, it is meaningful to plot the data for different products in one plot, see Figure \ref{fig:2WVPoverTimeScaled}. Having standardized the observations it should be noted that even in an ideal world the standardized observations are no longer independent to each other. The fewer observations there are per product, the higher the dependence of the standardized measurements for different production lots, as the correlation matrix of the standardized values:
\begin{equation}
\left(\begin{matrix}
	\frac{n-1}{n} & \frac{-1}{n} & \frac{-1}{n} & \dots & \frac{-1}{n}\\
	\frac{-1}{n} & \frac{n-1}{n} & -\frac{-1}{n} & \dots & \frac{-1}{n}\\
	\vdots &  & \ddots &  & \vdots\\
	\frac{-1}{n} & \frac{-1}{n} & -\frac{-1}{n} & \dots & \frac{n-1}{n}\\
\end{matrix} \right) * \sigma,
\end{equation}
with $\sigma$ being the standard deviation of the random variables.
This is especially notable, as the majority of products have a rather low number of production lots per year.

\subsection{Control charts}

After transforming the data, now it is possible to apply control charting methods to these data jointly. There is a wide field of new developments for control charting going in directions as multivariate control charts, non-parametric versions, control charts using supervised and unsupervised learning methods, control charts for functional data and many more topics. For an overview of these topics, please refer to \cite{Qiu2014IntroToSPC}.
Here still there are rather simple methods applied. 
Two types of control charts are used here: The one is an Individual Range chart (IR chart) and the other one is an Exponentially Weighted Moving Average chart (EWMA chart). 

An IR chart consists of two plots for the data, one plotting the individual measurement data, and one plotting the moving range between each pair of neighbouring points. The x-axis should be sorted in the order of productions. For the individual measurement plot, upper and lower control limits are added calculated based on using the moving range data as an estimator for the process variability. The corresponding formulas can be found at page 268 of \cite{MontgomeryIntroSPC2001}. Also for the moving range plot, an upper control limit is calculated. All control limits are calculated so that it holds that a stable process is producing observations, which are inside the control limits with a 3-sigma probability, i.e. the probability of rating a stable process as out of control is lower than $0.0027 \% $. As for all control charts, the general rule to define a process being out of control is, if there is at least one data point outside the control limits. For an example, please see Figure \ref{fig:4SimulatedDataIRchart}.

The aim of an EWMA chart is to increase the ability of the control chart to detect smaller changes in the process. In contrast to the IR control chart it consists of only one chart for the moving averages. The individual data points are transformed in the following way:

\begin{equation}
	z_i = \lambda y_i + (1 - \lambda) z_{i - 1},
\end{equation}
with $z_0 := \bar{y}$. The constant $\lambda \in (0, 1)$ is controlling how strong the smoothing is and needs to be set by the user. Here it is used $\lambda = 0.2$ throughout this paper, which is the default value in the software used  \cite{JMP12_2015QualityAndProcess}. The data points $z_i$ are plotted and control limits are calculated, see e.g. \cite{MontgomeryIntroSPC2001} page 434 for details. The EWMA control chart is often also referred to as a memory control chart, as it incorporates the process history into the current evaluation of the process. 

Motivation for using the IR chart is that the standard approach is currently to test each work order once. So the IR chart conceptionally fits well to the data at hand. Also, IR charts are a tool which is well understood by engineers and other job functions related to quality. If the data structure would be different, other charts like Shewart charts or three way charts could be used. 
Motivation for using the EWMA chart is to apply smoothing to the data as the data used here often show a high variability compared to relevant signals. 

A measure to describe the performance of a control chart is the Average Run Length (ARL), which is defined as the average number of observations for a control chart to produce an out of control observation, starting from the last out of control signal. It can be shown, that e.g. for conventional Shewart charts, the ARL of an in-control process is $\frac{1}{p}$, with $p$ being the probability belonging to a sigma level, see \cite{MontgomeryIntroSPC2001}, p. 199. Most of the time, with $p = 0.0027, ARL \approx 370$.
The $ARL$ is often looked at either under the assumption of an in-control process ($ARL_0$) or under the assumption of an out-of-control process ($ARL_1$). While $ARL_0$ should be as large as possible, it is attractive to have $ARL_1$ as small as possible. As both points are not perfectly achievable at the same time, a compromise has to be taken. In this article, $ARL_0$ and $ARL_1$ will be used to compare the different options of standardizations for control charts. These comparisons are done based on simulations. 

 
\section{Simulation study}
The aim of the simulation study performed here is twofold: First to identify, which standardization performs best for a stable process and secondly to study, how the performance of the chosen method is to detect an out of control status. 

\subsection{Average run length comparison under a stable process}
In the comparison, the arithmetic mean and the median are used as estimators for the data centre. In order to estimate the variability, the (sample) standard deviation, robust standard deviation, interquartile distance and the median absolute deviation are used. 
Hence there are 8 combinations of standardization methods.
As simulation scenario, the data set of 1 year of production data at Gore is used, i.e. from these data the (cleaned) averages and standard deviations for each product produced in one year are estimated. Then the amount and  chronological order of the products produced are used for simulating random numbers generated by using the estimated average and standard deviation per product. 
As the real data tend to produce outliers, for 1\% of the data, an outlier process is added to the simulated data with a much bigger standard deviation:
	\begin{equation}
	y_{ij} = \mu_{i} + e_{ij} + B * 25 * e_{ij,ol},
	\end{equation}
with $var(e_ij) = \sigma_i^2$, $B$ being a $\mathbb{B}(1, 0.01)$ distributed Bernoulli random variable and $var(e_{ij, ol}) = 25 \sigma_i^2$. 

Including an outlier process for a control chart simulation assuming a stable process seems to be illogical at first sight. However, in the authors experience measurement processes often have to deal with outliers, which do not necessarily relate to a manufacturing problem. Furthermore, in this special case of control charting, we are not interested to detect single out of control datapoints, but to detect, if there are persisting root causes for unstable processes. For reference, data under a stable process without outliers are generated as well.
The data of one simulation run is used to calculate the different variants of summary statistics and control charts.



\begin{table}
\centering
\begin{tabular}{|c|c|r|r|r|r|}
\hline
Outliers &	 & IQR & MAD & robust std dev & std dev\\
\hline
excl. &Mean & 190.8 & 132.4 & 463.6 & 777.9 \\	
\hline 
excl.& Median & 144.5 & 117.3 & 391.8 & 676.2 \\
\hline 
incl. &Mean & 110.9 & 76.4 & 157.7 & 174.5 \\
\hline 
incl. &Median & 109.4 & 78.8 &151.2  & 159.9 \\
\hline 
\end{tabular}
\caption{$ARL_0$ simulation comparison assuming a stable process and an IR Chart.}
\label{Tab1:ARL0simreults}
\end{table}

The results are shown in Table \ref{Tab1:ARL0simreults}. There is a clear indication, that there are big differences in terms of average run length. As expected, the ARL of the simulation runs including outliers is lower than for simulation runs without outliers. One surprising point on these data is that the MAD is providing a very low $ARL_0$ under a stable process. Analysing this behaviour it turns out, that especially for products with very few production lots, there is a high likelihood that the MAD will result in a value very close to zero. Using these values in the standardization results then in many very high/low values, which are detected as out of control signal in the control chart.

As stated above, in theory the $ARL_0$ should be around 370. Interestingly, exlcuding outliers, using the robust standard deviation in combination with the median is coming close to this value. The standard deviation, combined both with the median and the mean, delivers much higher agerage run length. This is an indication, that the induced correlation structure has an influence on the performance of the run length of the IR chart. Including the outlier process is reducing the $ARL_0$ by more than 50\%. Out of this analysis, the MAD and interquartile range are unattractive in terms of delivering too many false alarms.

\subsection{Average run length comparison under an unstable process}
For studying the behaviour under an unstable process the same data structure and model equation as assumed in the previous section is used.
The control limits are estimated from a phase of a stable process. 
Then in a second phase, the process is disturbed by different kinds of signals. 
The data over a year is split into $3/4$ for phase I and $1/4$ for phase II of the control chart.
Three different specific root causes are selected, i.e. it is not assumed that the overall process is out of control (as this would be very unlikely for the real process) but it is assumed that e.g. a specific raw material used in several but not all products is producing failures. 

Then for this root cause, there is a (positive, non-random) shift with $2 \sigma_i, 4\sigma_i, 6 \sigma_i$ and $8 \sigma_i$ added to the random numbers for each of the products, which are affected by that root cause. The values $(2, 4, 6, 8) \sigma$ are to some degree arbitrary and have been chosen to represent small and large shifts in average performance.

The three root causes picked are selected based on their occurrence in the data table. A tooling part, which has been used in not too many products (67 production lots, root cause A),
a raw material, which is used in a medium number of products (110 production lots, root cause B)
and another raw material which is used in a high amount of products (511 production lots, root cause C)
have been selected. 
As above, this process is repeated $n_{sim} = 500$ times with different seeds.

\begin{figure}
\centering
  \includegraphics[width=.75\linewidth]{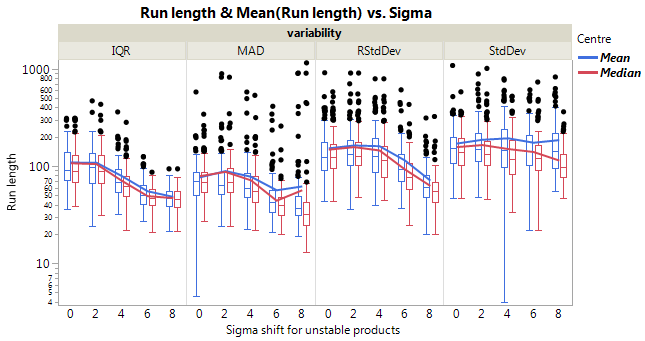}
  \includegraphics[width=.75\linewidth]{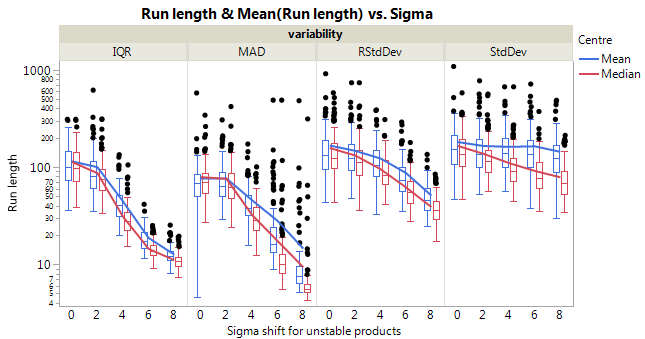}
  \includegraphics[width=.75\linewidth]{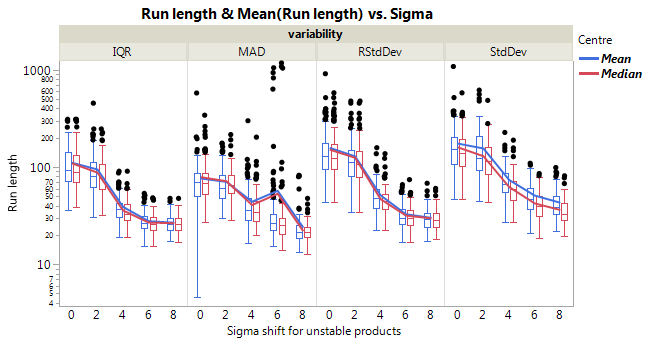}
\caption{Simulation results for an unstable process according to the three selected root causes and with different magnitudes of out of control signals. The y-axis is showing the simulated $ARL_1$ data on a log scale.}
\label{fig:2ComparisonUnstableProcess}
\end{figure}

\begin{table}
\centering
\begin{tabular}{ | r | r | r | c | c | c | c | }
	\hline
	root cause & Sigma & Centre &  IQR & MAD & RStdDev & StdDev \\ \hline 
	\multirow{10}{*}{A} & 0 & Mean & 110.9 & 76.4 & 157.6 & 174.5 \\ \cline{2-7}
	 & 0 & Median & 109.4 & 78.8 & 151.2 & 159.9 \\ \cline{2-7}
	 & 2 & Mean & 108.4 & 88.9 & 162.8 & 186.6 \\ \cline{2-7}
	 & 2 & Median & 104.8 & 87.2 & 156.8 & 164.6 \\ \cline{2-7}
	 & 4 & Mean & 78.7 & 77.7 & 160.2 & 193.1 \\ \cline{2-7}
	 & 4 & Median & 70.4 & 71.3 & 144.8 & 149.9 \\ \cline{2-7}
	 & 6 & Mean & 55.1 & 56.4 & 114.2 & 174.1 \\ \cline{2-7}
	 & 6 & Median & 49.1 & 44.1 & 92.4 & 139.6 \\ \cline{2-7}
	 & 8 & Mean & 48.2 & 61.8 & 70.9 & 185.1 \\ \cline{2-7}
	 & 8 & Median & 46.8 & 56.5 & 62.5 & 114 \\ \hline
	\multirow{10}{*}{B} & 0 & Mean & 110.9 & 76.4 & 157.7 & 174.5 \\ \cline{2-7}
	 & 0 & Median & 109.4 & 78.8 & 151.2 & 159.9 \\ \cline{2-7}
	 & 2 & Mean & 99.7 & 76.4 & 147 & 164.9 \\ \cline{2-7}
	 & 2 & Median & 86.6 & 75.8 & 131.8 & 137 \\ \cline{2-7}
	 & 4 & Mean & 44.3 & 45.6 & 122.7 & 161.6 \\ \cline{2-7}
	 & 4 & Median & 30.9 & 32.2 & 95.2 & 110.1 \\ \cline{2-7}
	 & 6 & Mean & 18.6 & 27.9 & 87 & 163.7 \\ \cline{2-7}
	 & 6 & Median & 14.4 & 17.3 & 61.9 & 91.4 \\ \cline{2-7}
	 & 8 & Mean & 12.6 & 14.5 & 51 & 143.5 \\ \cline{2-7}
	 & 8 & Median & 11.1 & 9.3 & 38.8 & 78.5 \\ \hline
	\multirow{10}{*}{C} & 0 & Mean & 110.9 & 76.4 & 157.7 & 174.5 \\ \cline{2-7}
	 & 0 & Median & 109.4 & 78.8 & 151.2 & 159.9 \\ \cline{2-7}
	 & 2 & Mean & 94.5 & 70.4 & 128.9 & 155.5 \\ \cline{2-7}
	 & 2 & Median & 87.7 & 71.5 & 124.6 & 129.4 \\ \cline{2-7}
	 & 4 & Mean & 39 & 44.1 & 51.2 & 73.8 \\ \cline{2-7}
	 & 4 & Median & 35.7 & 40.7 & 46.8 & 62.4 \\ \cline{2-7}
	 & 6 & Mean & 27.8 & 56.9 & 32.7 & 51 \\ \cline{2-7}
	 & 6 & Median & 26.9 & 53.1 & 31.7 & 42.3 \\ \cline{2-7}
	 & 8 & Mean & 26.8 & 23.8 & 30 & 42.9 \\ \cline{2-7}
	 & 8 & Median & 26.4 & 21.9 & 29.7 & 36.2 \\ \hline
\end{tabular}
\caption{$ARL_1$ simulation comparison assuming a unstable process and an IR Chart.}
\label{tab:2ARL1unstableprocess}
\end{table}

The results are shown in Figure \ref{fig:2ComparisonUnstableProcess} and Table \ref{tab:2ARL1unstableprocess}. Please note, that the results for a stable process are included in this table as well for each root cause for having a better comparibility (Sigma $= 0$). From these simulations it turns out that the robust standard deviation is attractive, as it provides a realistc $ARL_0$ and at the same time is sensible for out of control situations. For the robust standard deviation, the difference between the median and the mean ist not very high with a slightly better $ARL_1$ for the median. Interestingly, the standard deviation for root cause A and B delivers substantially higher $ARL_1$ than the robust standard deviation. As well, for the standard deviation, there is a difference in $ARL_1$ visible between mean and median, which is not as high for the robust standard deviation.  

Hence, the transformation which will be applied in the following is the robust standard deviation (using an M-estimator) combined with the median.

\subsection{Root cause analysis for an unstable process}

If it is detected by the above methodology, that the process is out of control, a next step is to apply data mining techniques to identify potential root causes. 
Important to note here is that even if an input factor is identified by data mining methods to be a root cause, this does not directly imply causality. As the data used for the analysis is an observational data set with e.g. a lot of correlations, a final assessment about root causes can not be made.
However, even getting suggestions for possible root causes can help a lot in a trouble shooting situation in order to prioritize and down select next actions.
Here no general introduction to the topic of data mining should be given, but rather an outlook, how such kind of data can be analysed. 
For good introductions into machine learning/data mining, interested readers can refer to \cite{HasTibFri01elem} or \cite{Wittenetal2011DataMining}.

One first step could be to apply an ordinary least squares model. However, this is not done here. One reason is, that failures often do occur suddenly and not as a long term trend. Hence applying a methodology which can separate shifts is attractive. As long as the date of change in quality is unknown, it would be not as straight forward to interpret the result. A method which can naturally deal with sudden shifts in data are partition trees, in contrast to ordinary least squares. 
\begin{figure}[ht]
  \centering
  \includegraphics[width=\textwidth]{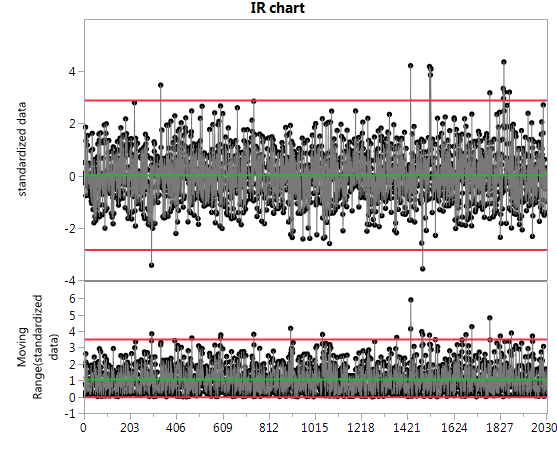}
\caption{IR chart of the standardized observations for a simulated data set.}
\label{fig:4SimulatedDataIRchart}
\end{figure}

\begin{figure}[ht]
  \centering
  \includegraphics[width=\textwidth]{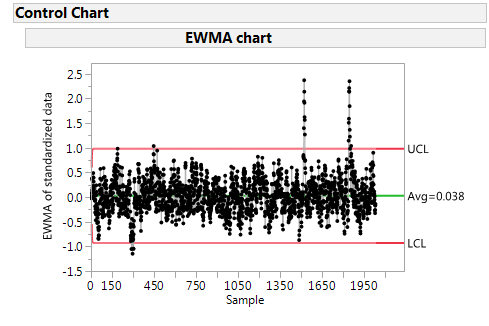}
\caption{IR chart of the standardized observations for a simulated data set.}
\label{fig:5SimulatedDataEWMAchart}
\end{figure}

\begin{figure}[ht]
  \centering
  \includegraphics[width=\textwidth]{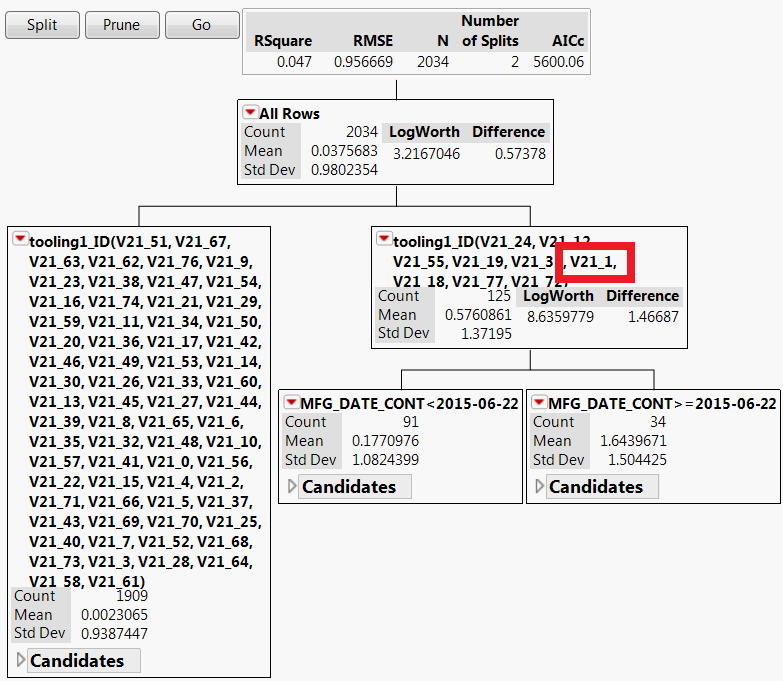}
\caption{Partition tree of the standardized observations for a simulated data set.}
\label{fig:6.1SimulatedDatapartitiontree}
\end{figure}
Also different kinds of neural networks are proven to be very effective in dealing with observational data as well as bootstrap forest.
However, although e.g. neural networks and bootstrap forests are often described as being better predictors here a partition tree is applied mainly because with a partition tree, precise suggestions (according to the tree structure) about potential root causes can be derived. Although for any kind or predictive model it is still possible to get an estimate about the importance of a factor, it is often not so straight forward to say which levels of that factor differentiate to other levels. Partition trees on the other hand can directly show which levels have been separated into different branches of the tree.
Here the transformed data points $y^{(r)}_{ij}$ are used as output and all columns available for root cause analysis are used as predictors. In this case study there are 8 factors used in the partition tree: The manufacturing date, in order to identify, when a problem started and 7 columns describing raw materials, toolings, process line and an operator effect.

\section{Application}
\begin{figure}[ht]
	\centering
	\includegraphics[width=\textwidth]{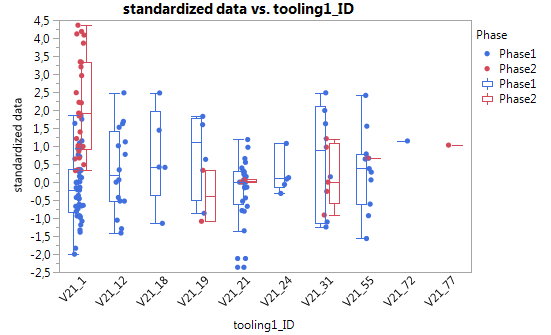}
	\caption{Visual analysis of standardized values versus different tools, splitted by the two time periods in the partition tree.}
	\label{fig:8GraphbuilderOOC}
\end{figure}

The complete flow of the method is presented on one of the simulated examples from section 4.2. However, as the simulation set-up is very close to the actual data sets data this gives a very realistic picture of how this framework is applied at W. L. Gore \& Associates. 
In the data set one component ($V21_1$) is simulated to have a deviation of 4.5 $\sigma$ of the corresponding products, starting from a certain point in time. 
As a first step the standardized values are calculated as described above followed by plotting the IR chart and the EWMA chart (see Figure \ref{fig:4SimulatedDataIRchart} and \ref{fig:5SimulatedDataEWMAchart}). 
For the IR chart and the EWMA chart there are two groups of out of control points in the last third of the chart, which can be seen more clearly in the EWMA chart. 
Applying a partition tree to these data reveals that there seems to be a specific reasoning for the out of control data points. 
Without specifying in advance, the partition tree detects tooling1ID as an important factor. Among the group of tools identified by the partition tree, there is also the one which was simulated to be out of control ($V21_1$). A next step in analysis could be use further splits in the partition tree or to visualize the tools on the right hand side of the first split, combined with the manufacturing date information. This is done in Figure \ref{fig:8GraphbuilderOOC}.
Here the standardized values are plotted against the corresponding tools. The manufacturing date is split into 2 phases as found by the partition tree. Clearly it is detectable, that for the tool $V21_1$ phase 2 data are deviating from the other data. 

Using these information in a real trouble shooting situation can be of substantial value, as it gives strong suggestions where to look for differences and root causes. This methodology is applied on a monthly basis for a number of different lab tests.	

\section{Summary}

In this application study, traditional methods for statistical process control are modified such that they can be applied in an environment, where one process is producing products with different characteristics. 
Key is to standardize the measurements values by product so that the data can be compared across different products and hence the process itself can be studied. Similar approaches are known as short run SPC in the literature. 
Picking estimators for the centre and the variability of product data it turns out that different estimators lead to quite different behaviour in terms of average run length, under a stable process as well under an unstable process. 
In order to have a compromise between a high $ARL_0$ and a sensitive $ARL_1$ the median and the robust standard deviation are suitable estimators for standardizing the data. One point of imprecision is that the induced correlation structure of the transformed data points is not taken into account explicitly. Based on the experience gained so far with this process, this is not leading to undesired behaviour, but it remains open to search for justification or modification, especially for small number of production lots per product. If it turns out that this leads to undesired results, more advanced approaches could be incorporated as e.g. described in \cite{Quesenberry1991Qchart} and \cite{CastilloMontgomery1994QchartEnahncement}. 

After standardization, either IR control charts or EWMA control charts are used, but also many other different kinds of control charts can be applied to the standardized data. This is potentially an area of future research.
Having identified an unstable process, it is important to identify as quickly as possible possible root causes. Therefore the standardized measurement data are combined with process data e.g. about raw material types and tooling information. 
Applying data mining techniques as partition trees enables a user to identify potential root causes. This can save substantial time and money in problem solving situations and hence is, in combination with control charting, a highly valuable pair of methods for assuring and re-establishing stable processes.
This method is applied already for approximately one year at W. L. Gore \& Associates and has proven to be very useful for process control. In case of interest, the data sets presented in this data set can be requested from the author. 


\bibliographystyle{abbrv}

\end{document}